\documentclass[aps,prl,twocolumn,showpacs,superscriptaddress]{revtex4}
\usepackage{amsmath,amssymb,graphicx,color}
\usepackage{textcomp}
\usepackage{ulem}

\begin{document}

%\preprint{}

\title{Lift and down-gradient shear-induced diffusion in Red Blood Cell suspensions}


\author{Xavier Grandchamp}
\author{Gwennou Coupier}
\author{Aparna Srivastav}\affiliation{Laboratoire Interdisciplinaire de Physique, CNRS - UMR 5588, Universit\'{e} Grenoble I, B.P. 87, 38402 St Martin d'H\`{e}res Cedex, France}

\author{Christophe Minetti}\affiliation{Microgravity Research Center, Universit\'{e} Libre de Bruxelles, 50 av. F. D. Roosevelt, B-1050 Brussels, Belgium}
\author{Thomas Podgorski}\email{thomas.podgorski@ujf-grenoble.fr}

%\thanks{}
\affiliation{Laboratoire Interdisciplinaire de Physique, CNRS - UMR 5588, Universit\'{e} Grenoble I, B.P. 87, 38402 St Martin d'H\`{e}res Cedex, France}

\date{\today}

\begin{abstract} 
The distribution of Red Blood Cells in a confined channel flow is inhomogeneous and shows a marked depletion near the walls due to a competition between migration away from the walls and shear-induced diffusion resulting from  interactions between particles. We investigated the lift of RBCs in a shear flow near a wall and measured a significant lift velocity despite the tumbling motion of cells. We also provide values for the collective and anisotropic shear-induced diffusion of a cloud of RBCs, both in the direction of shear and in the direction of vorticity. A generic down-gradient subdiffusion characterized by an exponent $1/3$ is highlighted.
\end{abstract}


\maketitle

Blood is a dense suspension of deformable cells, mainly red blood cells (RBCs), making it a complex fluid from a rheological viewpoint, and leading to complex flow patterns in the microcirculation where the diameter of blood vessels becomes comparable to cell size.

In his pioneering work, Poiseuille revealed that blood flow in arterioles and venules features a RBC-free plasma layer near the vessel wall \cite{sutera93,poiseuille1835}. The lubrication effect of this depleted layer leads to  the F{\aa}hr{\ae}us-Lindquist effect, a decrease of the apparent viscosity of blood in small vessels when their diameters become comparable to cell size ($d<500$\,\textmu m) \cite{popel05} .

A classic result in low-Reynolds number hydrodynamics --- relevant to blood flow in arterioles and venules \cite{goldsmith71,fung93} --- is that migration of spherical particles transversally to flow direction is prohibited by the linearity and flow-reversal symmetry of the Stokes equation \cite{bretherton62}. However, the deformability or the non-sphericity of RBCs allow a symmetry breaking that may lead to transverse migration, be it due to interactions with walls or neighboring cells.

In a shear flow near a wall, lipid vesicles experience a lift force that pushes them away \cite{olla97, cantat99,abkarian02, callens08}, at least when they are in a tank-treading regime, with steady inclination angle. A straightforward question arises: how do RBCs, that are usually in a tumbling regime \cite{goldsmith71,goldsmith72} and explore all angles, still experience a non-zero average lift force ? While many numerical studies have tried to reproduce this behavior \cite{bagchi07, secomb07,shi10,li10, kumar12,hariprasad12}, experimental data on this basic mechanism are rare \cite{goldsmith71} or focused on RBCs artificially placed in the tank-treading regime \cite{geislinger12}.

This migration of blood cells forms the physical basis of the formation of a depleted layer near vessel walls in the microcirculation. However this phenomenon alone cannot explain the complexity of flow patterns observed in the microvasculature, where redistribution processes are indeed very frequent since bifurcations are met every 20 vessel radii \cite{risser09}. In physiological conditions, blood is a very concentrated suspension with a hematocrit up to 50\%, in which the hydrodynamic interactions between cells play a decisive role. 
The interactions between two bodies in flow is a fundamental question in the framework of suspension dynamics and rheology, even in rather dilute suspensions \cite{batchelor72a,zinchenko84}. Unlike smooth and spherical particles, rough spheres \cite{dacunha96} and deformable particles such as drops, bubbles, capsules or vesicles \cite{wijngaarden76, loewenberg97, lac07, kantsler08,gires12}, are irreversibly shifted after interaction.

The cumulative effect of these hydrodynamic interactions
is a non-linear and anisotropic shear-induced diffusion (SID) \cite{loewenberg97,dacunha96}.
The consequences of this SID are twofold: repeated collisions of one blood cell with the others lead to a random walk which may be important for mixing properties of blood flows (self diffusion), 
and a redistribution of concentration inhomogeneities, that balances lift forces (collective or down-gradient diffusion). The coefficients characterizing both phenomena are \textit{a priori} different \cite{dacunha96}. 
Investigations of the random walk of RBCs in concentrated suspensions \cite{goldsmith71,goldsmith79,cha01} provided values of the dispersion coefficient two orders of magnitude higher than Brownian diffusivity. These studies were complemented by cell tracking experiments in quasi 2D flow \cite{Higgins09}.
All these studies and most numerical works \cite{zhao11,tan12} focus on the self diffusivity, however there has been no experimental quantification of  down-gradient diffusion in non-homogeneous suspensions
of RBCs and whatever the considered particles, experimental characterizations of down-gradient diffusion are scarce \cite{hudson03,rusconi08}.

We report on an experimental study on the lift of diluted  RBCs in shear flow near a wall and show that tumbling RBCs follow the same scaling laws as tank-treading vesicles \cite{olla97,callens08}, capsules \cite{pranay12}  and drops \cite{hudson03}. In a different experiment, the collective SID of RBCs was investigated, providing values of the diffusivity in the vorticity direction and in the direction of shear.

\paragraph{Lift of RBCs in shear flow near a wall.}--- In order to avoid screening  by sedimentation, measurements were performed in microgravity in CNES and ESA parabolic flight campaigns. 
The procedure and experimental setup are detailed in Ref. \cite{callens08}. We use a Couette shear flow chamber with two parallel glass discs, with a gap of 170\,\textmu m. The 3D positions of the  RBCs 
which initially lie on the bottom disc are captured by digital holographic microscopy \cite{Dubois06_2,dubois06_1}.

Blood was collected from healthy donors and washed twice in Phosphate Buffer Saline (PBS) and Bovine Serum Albumin (BSA). After gentle centrifugation, RBCs were dispersed into different fluids: (PBS)+(BSA) alone or combined with a mixture of 1$\%$ dextran of molecular weight $1.5\times10^{4}$ + $n\%$ dextran of molecular weight $2\times10^{6}$, $n=3,4,5$. Corresponding viscosities are respectively 1.4, 6.1, 9.3 and 13\,mPa.s ($T \approx 21^{\circ}$C). 

For a neutrally buoyant ellipsoidal lipid
vesicle, the theoretical drift velocity is given by \cite{olla97}: $\dot{z}=U\dot{\gamma}R^{3}/z^{2}$, where $R$ is a particle characteristic size, $\dot{\gamma}$ the shear rate, and $z$ the distance to the wall. \textit{U} is a dimensionless drift velocity which depends on vesicle shape, and on the inner and outer fluid viscosities. It yields the following scaling:  
\begin{equation}
z ^{3}=3 UR^3\dot{\gamma}t + z_0^3
\label{drift}
\end{equation}

\begin{figure}
\includegraphics[width=\columnwidth]{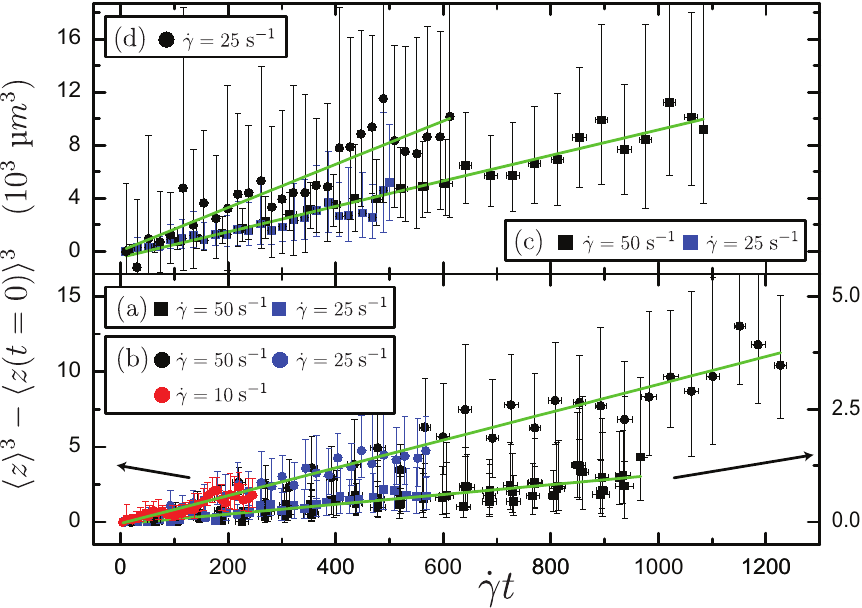}
\caption{\label{driftvelocity}(Color online) RBC - wall distance $\langle z \rangle ^{3}$  vs.  $\dot{\gamma} t$ for different outer viscosities: (a) 1.4\,mPa.s; (b) 6.1\,mPa.s; (c) 9.3\,mPa.s; (d) 13\,mPa.s. Full lines indicate fit to Eq. \ref{drift}, with $U R^3= 0.36, 3.1, 3.2, 5.4$\,\textmu m$^3$, respectively.}
\end{figure}

The evolution of the mean transverse position $\langle z \rangle ^{3}$ is presented as a function of $\dot{\gamma}t$ in Fig.\ref{driftvelocity}.
For a given external solution all results fall on the same straight line in agreement with Eq.\ref{drift}. By symmetry, a tumbling rigid object should not migrate on average \cite{olla97}. The non zero lift 
suggests that RBC deformability allows symmetry breaking: it is stretched when oriented in the direction of the elongational component of the flow, while it is compressed when orthogonal, resulting in an averaged asymmetric shape, leading to a migration law similar to the one known for a fixed shape and orientation. 
By increasing the external viscosity, the stresses on the RBC membrane are higher and lead to increased deformation \cite{goldsmith72} which in turn enhances the lift.
RBCs in the 13\,mPa.s solution have viscosity ratio close to 0.3 \cite{koter90} and are probably very close to tank-treading regime \cite{fisher07}, in which case Olla predicts a migration with comparable $U R^3=6.4$\,\textmu m$^3$  for a vesicle of similar (but fixed) shape \cite{olla97}, with long axis equal to 7.2\,\textmu m, the mean diameter  of a RBC \cite{turgeon05}. Assuming $\dot{x}=\dot{\gamma}y$, we find that in  physiological conditions, a RBC will migrate by 8\,\textmu m while travelling 1 cm, a result in good agreement with the pioneering result of 4\,\textmu m drift  in Poiseuille flow by Goldsmith \cite{goldsmith71}. 

\begin{figure*}[t!]
Ê\includegraphics[width=2\columnwidth]{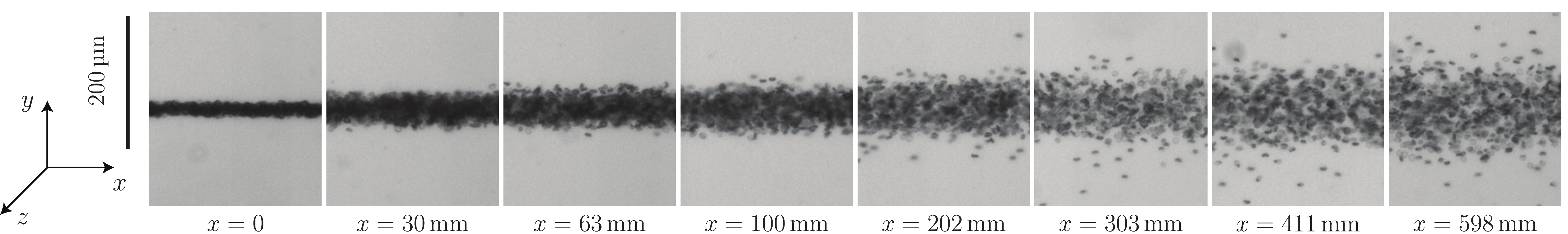}
\caption{\label{fig:exemple}Example of RBC diffusion in a flat channel. Initial mean volume fraction is around 15\%.}
\end{figure*}

\paragraph{Shear-induced diffusion in channel flow}--- The SID of a RBC suspension was studied in standard polydimethylsiloxane microfluidic chips. 
Thanks to a flow-focusing device, a thin layer of RBC suspension is produced in a rectangular channel  where a buffer (PBS solution) flows in the $x$ direction, the gravity direction. The RBC cloud is pinched at the entrance in the $y$ direction and the direction of observation is $z$, allowing to record the evolution of the RBC cloud in the $(x,y)$ plane (Fig. \ref{fig:exemple}). We restrict the study to moderate shear rate values bounded by shear rate at the edge $\dot{\gamma}_{\max}\le 340$\,s$^{-1}$ and therefore comparable to physiological ones \cite{goldsmith71}. 

We first focus on two channels with high aspect ratios: width $2d\times$height $2h=491\times$53\,\textmu m$^{2}$ and 497$\times$101\,\textmu m$^{2}$. Thus, the velocity profile is parabolic across \textit{Oz} and almost flat in the \textit{Oy} direction. In this case, due to the strong shear in the $z$ direction, the concentration tends to homogenize quickly due to diffusion in the plane of shear, while diffusion in the vorticity $y$ direction leads to the observed widening (Fig. \ref{fig:exemple}). 

From microscopic images taken with a long exposure time, a calibration process based on the Beer-Lambert law relates the grey intensity to the local concentration profile $\phi(x,y)$. The evolution of the concentration profile along $x$ is directly related to the diffusion process through the following advection-diffusion equation \cite{suppl}:

\begin{equation}
\langle u\rangle\frac{\partial \phi}{\partial x}=\frac{\partial}{\partial y} \left( f_3 R^2 \dot{\gamma}\phi \frac{\partial \phi}{\partial y} \right) = f_3 R^2 \langle \dot{\gamma}\rangle \frac{\partial}{\partial y} \left(  \phi \frac{\partial \phi}{\partial y} \right),
\label{eq_flat}
\end{equation}
where $\langle.\rangle$ denotes the average over $z$.

Here we assume that the concentration of RBCs is homogeneous in the $z$ direction. We also suppose that the velocity $u$ is the one of an unperturbed Newtonian fluid \cite{white}. The diffusivity $D=f_3 R^2 \dot{\gamma}\phi$ is proportional to the frequency of pair interactions $\dot{\gamma}\phi$, a straightforward scaling for shear-induced diffusion due to pair interactions \cite{dacunha96,loewenberg97}. As in Ref. \cite{dacunha96}, we denote $f_3$ the dimensionless diffusivity in the vorticity direction.

\begin{figure}[t!]
Ê\includegraphics[width=\columnwidth]{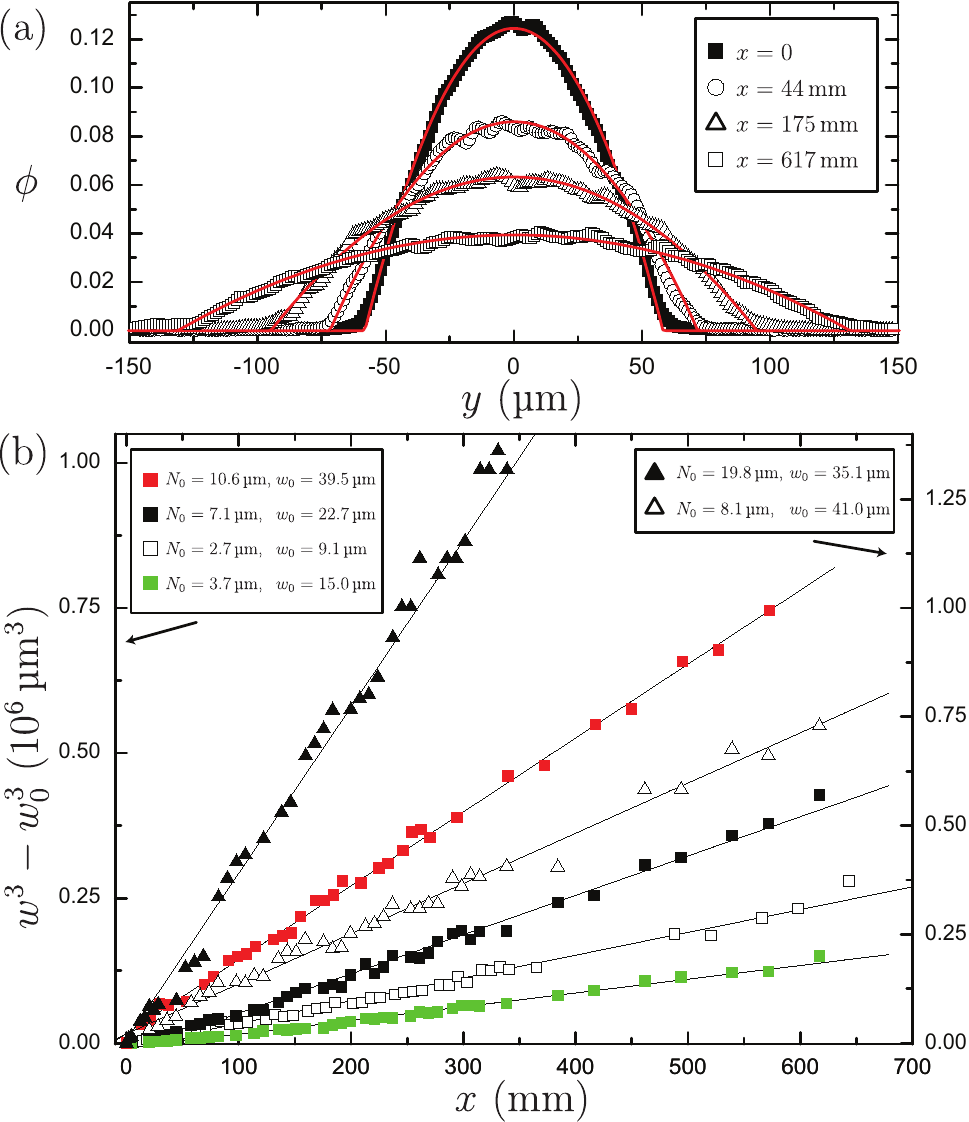}
\caption{\label{fig:w}(color online) (a): concentration profiles $\phi(x,y)$ in four sections of flat channel with $2h=53$\,\textmu m, for a cloud with $w_0=41.0$\,\textmu m and $N_0=8.1$\,\textmu m. Full lines show fits with parabolic profile. (b) Cloud half-width $w$ as a function of position $x$ along the channel for several initial conditions and for two different thicknesses (empty symbols, $2h=53$\,\textmu m, $\dot{\gamma}_{\max}=113$\,s$^{-1}$; full symbols, Ê$2h=101$\,\textmu m, $\dot{\gamma}_{\max}=211$\,s$^{-1}$). Full lines show linear fit for $w^3$ as suggested by Eq. \ref{w_flat}.}
\end{figure}

\begin{figure}[t!]
Ê\includegraphics[width=\columnwidth]{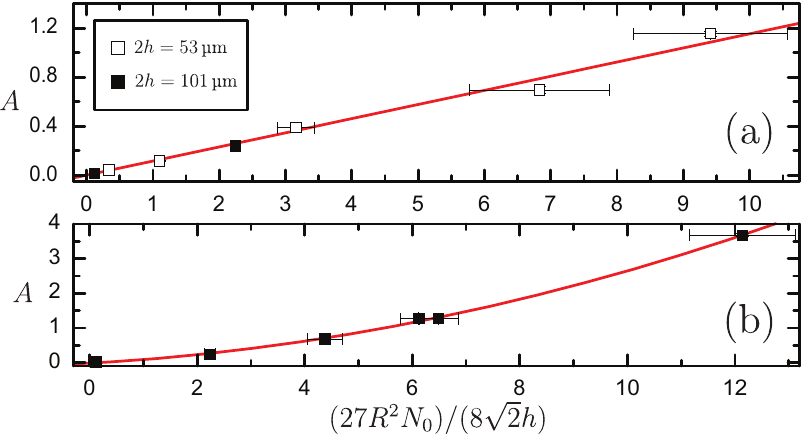}
\caption{\label{fig:f3quadra} Effective diffusion coefficient $A$ as a function of $N_0/h$ in  a flat channel. (a):  $2h=53$\,\textmu m  and $2h=101$\,\textmu m (data restricted to initial maximal concentration between 3 and 16\%,  resp. 2 and 12 \% and $N_0<3.7$\,\textmu m). Full line shows linear fit. (b):  $2h=101$\,\textmu m. Data are extended to initial maximal concentration 30\% Êand $N_0<19.8$\,\textmu m (curve with the largest slope in Fig. \ref{fig:w}). Full line shows quadratic fit	.}
\end{figure}

Rusconi and Stone made similar experiments with platelet like particles, with different initial and boundary conditions \cite{rusconi08}. They considered the spreading of a concentration step with fixed concentrations at each end and found a $x^{1/2}$ scaling for the diffusive front.  In our case, a peak of fixed area spreads and self-similar solutions exist under the condition of a widening with a $x^{1/3}$ scaling \cite{suppl}. The self-similar concentration profile is parabolic  and one finds the following relation for the expected half-width at half-height of the RBC cloud: 
\begin{equation}
w(x)=w_0 \left(1+A x/ w_0^3 \right)^{\frac{1}{3}},
\label{w_flat}
\end{equation}
where the initial peak has width $2w_0$, $A=\frac{27 f_3 R^2 N_0}{8 \sqrt{2} h}$ and $N_0=\int \phi(x,y) dy$ is the conserved number of particles \cite{suppl}. The scaling $w^3-w_0^3 = A x$ as well as the parabolic concentration profiles are nicely recovered in experiments for different $h$,  $w_0$ and $N_0$ (Fig. \ref{fig:w}), and the slope $A$ gives a direct measurement of $f_3$. Fig. \ref{fig:f3quadra}(a) shows that for all available data in the mid concentration range ($\phi<16\%$) $A$ is a linear function of $N_0/h$, giving a dimensionless diffusivity for RBCs $f_3=0.12\pm0.01$, with $2R=7.2$\,\textmu m, the mean diameter of a RBC. With similar choice for $R$, $f_3=6.9$ was found for very flat platelike particles \cite{rusconi08}. This discrepancy cannot be related to the deformability of RBCs:  self diffusivity of hardened cells has been shown to be of the same order as the one of normal cells \cite{Higgins09}. However, both discoidal particles are tumbling, thus the effective occupied volume is much larger. Replacing $\phi$ by $\phi V_e/V$, where $V$ is the particle volume and $V_{e}=4\pi R^3/3$ this effective volume, we find $f_3=0.05$ for RBCs and $f_3=0.18$ for platelike particles. These values are now comparable. The remaining difference can be attributed to the details of the hydrodynamic interactions. For instance, in Ref.  \cite{dacunha96}, $f_3$ varies from 0 to 0.03 for rough spheres with minimal separation going from 0 to $0.08R$.

At higher volume fractions of RBCs ($\phi<30\%$ in the experiments), the diffusion should not be a consequence of pair-wise interactions only, since a RBC interacts with multiple neighbors, and interactions between at least three bodies should also be considered. These interactions lead to an additional term proportional to $\phi^2$ in the diffusion coefficient and to a different scaling, namely $x^{1/4}$ if these 3-body interactions were the dominant effect \cite{suppl}. However, the noise in the experimental data allows a rather good rescaling with exponents between $1/3$ and $1/4$. By forcing a $1/3$ exponent, one gets an effective diffusivity which increases more than linearly with concentration, showing the increasing importance of 3-body interactions in the diffusive process (Fig. \ref{fig:f3quadra}(b)).

\begin{figure}[t!]
Ê\includegraphics[width=\columnwidth]{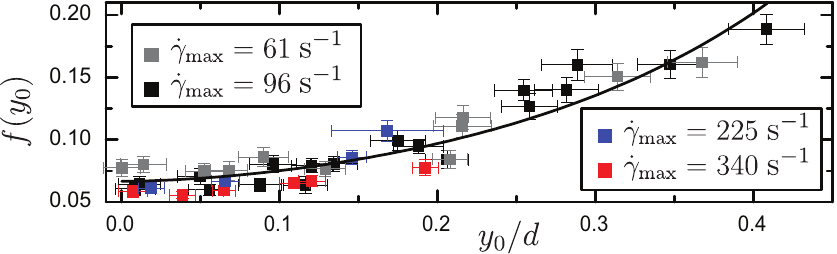}
\caption{\label{fig:f2f3}(color online) Effective diffusion coefficient $f(y_0)= 8 \sqrt{2}  A h /(27 R^2 N_0)$ in the  190$\times$99\,\textmu m$^{2}$ channel  as a function of lateral position $y_0$. $f$ is expected to converge to $f_3$ when $y_0\to 0$. Full line shows fit to theory.}
\end{figure}

An interesting potentiality of this experimental device is the possibility to measure both diffusivities $f_2$ and $f_3$ corresponding to repulsion of interacting cells in the plane of shear and in the vorticity direction \cite{loewenberg97}. In a  channel with cross section 190$\times$99\,\textmu m$^{2}$, a nearly parabolic flow with gradients of velocity in both directions $y$ and $z$ was produced. The averaged concentration profile observed in the $z$ direction therefore widens due to hydrodynamic repulsion in the local shear and vorticity directions. By varying the initial position $y_0$ of the RBC stream, one can vary the weight of the $f_2$ and $f_3$ contributions, with a contribution of $f_3$ only for $y_0=0$ and an increasing contribution of $f_2$ as the stream is moved towards the channel edges.
With the additional simplification that in the $y$ direction all particles experience the velocities and shear of position $y_0$
(narrow cloud approximation), 
one gets an equation similar to Eq. \ref{w_flat}, with coefficient $A$ that now depends on $y_0$: 
$ A=\langle \frac{f_2 u_y^2+f_3  u_z^2}{(u_y^2+u_z^2)^{1/2}}\rangle\frac{9 N_0 R^2}{4\sqrt{2}\langle u(y_0) \rangle} $, where $u_i$ is the partial derivative of $u$ according to variable $i=y,z$ at position $(y_0,z)$  \cite{suppl}.  Considering narrow initial clouds ($w_0\simeq6$\,\textmu m) consistent with the above simplification, the scaling with exponent $1/3$ is confirmed by the experiments and the resulting effective diffusion coefficient $A$ increases with $y_0$ (Fig. \ref{fig:f2f3}). Within the experimental uncertainties, $A$ does not depend on the mean shear rate, though the RBC dynamics and the consequent interaction trajectories might be affected by the shear rate value \cite{goldsmith72,abkarian07,dupire10}. A fit of the data by the expected expression yields $f_3=0.07\pm0.01$ and $f_2=1.7\pm0.1$. The $f_3$  value is lower than the one previously found. Around $y_0=0$, shear intensity in the $y$ direction vanishes, so does collision rate and diffusion is expected to be similar to the one observed in the flat channel, and controlled by $f_3$. However, $w_0$ is finite, and the 
3D shear also controls the mean orientation of RBCs, therefore the detail and intensity of their interactions, and finally the resulting diffusion coefficient $f_3$ may be affected. As for drops \cite{loewenberg97} or rough spheres \cite{dacunha96}, $f_2$ is found to be larger than $f_3$. In the case of drops,  experiments of Ref. \cite{hudson03} show that $f_2\simeq 0.2$, which is comparable to our $f_2=0.77$ obtained after rescaling by the effective volume. Finally, these  down-gradient diffusion coefficients should be compared to the self diffusion coefficients. Self diffusion in the vorticity direction was studied in Ref. \cite{Higgins09}, but the diffusion coefficient 
in this very flat geometry (12\,\textmu m thick channel)  
is surprisingly found to be independent from concentration.
By lack of similar scaling, comparison is therefore not possible.  Self diffusion in the shear direction is characterized by $D_s/\dot{\gamma}$ of order 1\,\textmu m$^2$ for $\phi\simeq40\%$  \cite{goldsmith71,goldsmith79,cha01,Bishop02}. We find $f_2R^2 \phi\simeq 9$\,\textmu m$^2$, a consistent result since this 
down-gradient coefficient is  expected to be a few times larger: 6 times for rough spheres \cite{dacunha96} and 5 times for drops \cite{hudson03}.


\paragraph{Conclusion.}--- Our quantitative investigation of the migration of tumbling RBCs in shear flow shows that wall-induced lift follows the same scaling law as particles with fixed orientation (e.g. tank-treading vesicles), and a significant  amplitude has been measured. The necessary symmetry breaking is made possible by RBC deformability. In blood vessels, lift is balanced by shear-induced diffusion. The spreading of a stream of blood cells in channel flow is characterized by a sub-diffusive behavior with exponent $1/3$, a phenomenon expected to be generic to systems where advected particles undergo short range pairwise hydrodynamic interactions or collisions. For RBCs, this scaling still holds at significant local concentrations where multibody interactions have to be considered in other systems such as rigid beads. We provide previously unpublished values of the down-gradient shear induced diffusivities of RBCs, with a marked difference between the diffusivity $f_2$ in the direction of shear and $f_3$ in the vorticity direction.  This strong anisotropy should be explained by a detailed analysis of RBC collisions at the microscopic scale. Our study pertains to dilute to semi-dilute suspensions of RBCs for which the convective lift flux due to lift of isolated cells and the diffusive lift flux can be readily balanced to get the concentration profile of the suspension in channel flow. At higher hematocrits though, the screening effect on the lift due to other cells as well as the local rheology of the concentrated suspension which modifies the flow profile should be considered.

\begin{acknowledgments}
The authors would like to thank CNES and ESA for financial support and access to microgravity platforms (parabolic flights) and B. Polack from Grenoble Hospital (CHU) and TIMC Laboratory for fruitful scientific discussions and technical advice on blood manipulation. Blood from healthy donors was provided by Etablissement Fran\c{c}ais du Sang (EFS Grenoble). This work was also supported by the SSTC/ESA-PRODEX (Services Scientifiques Techniques et Culturels/European Space Agency - Programmes de D\'eveloppement d'exp\'eriences) Contract No. 90171.
\end{acknowledgments}

$\,$

\newpage

\begin{center}
\textbf{Supplemental Material for "Lift and down-gradient shear-induced diffusion in Red Blood Cell suspensions
}\end{center}

In this supplemental material, we derive the advection-diffusion equation in a 3D Poiseuille flow. The resulting simplified equation in the case of a distribution invariant in one direction is similar to the one obtained in a simple shear flow. We then discuss the existence of self-similar solutions for these equations, whose shape strongly depend on the initial conditions considered. The solution relevant to our problem is then detailed.

\section{Advection-diffusion in a Poiseuille flow}

We consider the stationnary experiment where an initial concentration $\phi(0,y,z)$ is continuously injected at the inlet $x=0$ of a rectangular channel of section $2d\times 2h$ in the $(y,z)$ plane. The origin for $(y,z)$ coordinates is at the channel center. We neglect diffusion in the flow direction compared to advection. The diffusive flux in a simple shear flow  $\vec{J}=-D \vec{\nabla} \phi$ is different whether concentration gradient is in the velocity gradient direction or in the vorticity direction. In the first case, the diffusion coefficient $D$ reads $D=f_2 R^2 |\dot{\gamma}|\phi$ and in the second case it reads $D= f_3 R^2 |\dot{\gamma}|\phi$, where $R$ is the typical particle size.  In particular, the diffusivity is proportional to the rate of collisions $|\dot{\gamma}|\phi$. We assume that the particles velocity is independent from the concentration and we will generally consider that they flow with the same velocity $u(y,z)$ as the fluid, which they do not perturb. Expressions for Stokes flow in rectangular duct can be found in Ref. \cite{white}.

The deformation tensor $G$ is obtained from the velocity $u(y,z)$:
\begin{equation}
G=\begin{pmatrix}0&u_{y}&u_{z}\\ 
0&0&0\\ 
0&0&0\\ 
\end{pmatrix} ,
\end{equation}

where $u_i=\frac{\partial u}{\partial i}$, $i=y,z$. Writing $\Gamma=\sqrt{u_y^2+u_z^2}$, we find that in an orthonormal basis $(\mathbf{u_1},\mathbf{u_2},\mathbf{u_3})$, $G$ is transformed into: 
\begin{equation}
\begin{pmatrix}0&0&0\\ 
0&0&\Gamma\\ 
0&0&0\\ 
\end{pmatrix}.
\end{equation}

Locally, the flow can be described as a simple shear flow with shear rate $\Gamma>0$, where  $\mathbf{u_2}=(1,0,0)$ is the flow direction and 
$\mathbf{u_3}=\frac{1}{\Gamma}(0,u_y,u_z)$ gives the local velocity gradient direction. $\mathbf{u_1}=\frac{1}{\Gamma}(0,-u_z,u_y)$ gives the vorticity direction,. 

If $(x_1,x_3)$ are the local coordinates in the $(\mathbf{u_1},\mathbf{u_3})$ frame, the diffusive flux is by definition \begin{equation}
\mathbf{J}=-R^2 \phi \Gamma (f_2 \frac{\partial \phi}{\partial x_3} \mathbf{u_3}+f_3 \frac{\partial \phi}{\partial x_1} \mathbf{u_1}).
\end{equation}

The stationary advection-diffusion equation is:

\begin{equation}
u(y,z) \frac{\partial \phi}{\partial x}=-\frac{\partial}{\partial y}J_y-\frac{\partial}{\partial z}J_z,
\end{equation}

where the $\mathbf{J}$ components $J_y$ and $J_z$ are  obtained from preceding equation by standard variable manipulation:
\begin{eqnarray}
J_y&=& -R^2  \Gamma^{-1} \phi\Big[f_3  (\frac{\partial \phi}{\partial y}  u_{z}^2-\frac{\partial \phi}{\partial z}  u_{y}  u_{z} )\\ \nonumber
&&\qquad \qquad+f_2    (\frac{\partial \phi}{\partial y}  u_{y}^2+\frac{\partial \phi}{\partial z}  u_{y}  u_{z} )\Big],\\
J_z&=&-R^2 \phi \Gamma^{-1} \big[f_3  (\frac{\partial \phi}{\partial z}  u_{y}^2-\frac{\partial \phi}{\partial y} u_{y}  u_{z})\\ \nonumber&&\qquad\qquad +f_2    (\frac{\partial \phi}{\partial z} u_{z}^2+\frac{\partial \phi}{\partial y}  u_{y}  u_{z})\big].
\end{eqnarray}

If the distribution is initially $z$-invariant, we expect that it will remain so; $J_z=0$ and $\frac{\partial \phi}{\partial z}=0$ then implies:

\begin{equation}
 u(y,z) \frac{\partial \phi}{\partial x}=R^2 f_3  \frac{\partial}{\partial y}(u_{z}^2  {\Gamma}^{-1} \phi \frac{\partial \phi}{\partial y}  )+R^2 f_2 \frac{\partial}{\partial y}( u_{y}^2 {\Gamma}^{-1}\phi \frac{\partial \phi}{\partial y}  )
 \end{equation}

If the channel is flat, that is $d\to\infty$, the flow is parabolic in the $z$ direction, and $u_y=0$ so with no further assumption, we get: 

\begin{equation}
 u(z) \frac{\partial \phi}{\partial x}=R^2 f_3 |u_{z}|   \frac{\partial}{\partial y}(\phi \frac{\partial \phi}{\partial y}  )
 \end{equation}

For a more general section, if we consider a narrow cloud centered on $y_0$, we can make the simplification consisting in considering only the leading order in the development of $v$ and $u_y$ around $y_0$. The first one is constant, and the second one is constant if $y_0\ne0$ or of order 1 if $y_0=0$. Forgetting about this latter very specific case, we find: 

\begin{equation}
u(y_0,z) \frac{\partial \phi}{\partial x}= R^2 \,\frac{f_2 u_y^2+f_3  u_z^2}{\sqrt{u_y^2+u_z^2}}  \,\frac{\partial}{\partial y} (\phi \frac{\partial\phi}{\partial y} ). \label{eq:diffintegrated}
\end{equation}

Note that $u_y$ and $u_z$ are also functions of $y_0$. If we integrate in the $z$ direction, we get equation:

\begin{equation}
 \frac{\partial \phi}{\partial x}=\lambda(y_0) \frac{\partial}{\partial y} (\phi \frac{\partial\phi}{\partial y} ), \label{eq:eqdelta00}
\end{equation}

where  \begin{equation}\lambda(y_0)=R^2 \frac{ \int \Big( f_2\frac{u_y^2}{\sqrt{u_y^2+u_z^2}}  +f_3\frac{u_z^2}{\sqrt{u_y^2+u_z^2}} \Big) dz}{\int u(y_0,z) dz}\end{equation}
will contribute directly to the effective diffusion coefficient measured when considering the widening of a cloud of particles. 
For the flat channel, $\lambda(y_0)\equiv \lambda_{\mbox{flat}}= f_3\frac{3R^2}{2h}$ and in the general case where thickness $2h$ would not be too large compared to width $2d$, $\lim_{y_0\to 0}\lambda(y_0)\simeq \lambda_{\mbox{flat}}$.\\

Eq. \ref{eq:eqdelta00} is similar to the one that would be obtained in a shear chamber with flow in the $x$ direction. In such an experiment, the distribution is usually $x$-independent and its evolution with time can be considered. The shear plane is $xy$ and we have :

\begin{equation}
\frac{\partial \phi}{\partial t}=R^2 f_2  |\dot{\gamma}| \frac{\partial}{\partial y} ( \phi \frac{\partial\phi}{\partial y} )
\end{equation}

It is equation  \ref{eq:eqdelta00}  with $x$ replaced by time $t$ and $\lambda=f_2 R^2 |\dot{\gamma}| $.

\section{Self-similar solutions}

A slightly more general case than equation \ref{eq:eqdelta00} is considered for further discussion. In case of the usual Brownian diffusion, there is no $\phi$ term in the diffusion coefficient. In case of 3-body interaction, we have a $\phi^2$ term. For perfectly spherical particles, 2-body interactions do not lead to any diffusion in creeping flows, and 3-body interactions is the simplest mechanism leading to irreversibility and diffusion. If one of these kinds of diffusion is predominant, we  have equation:

\begin{align}
 \frac{\partial \phi}{\partial x} &= \frac{\partial}{\partial y} \big(\phi^m \frac{\partial \phi}{\partial y}\big), \label{eq:eqdiff}
\end{align}
where $m=0,1,2$ and $\phi$ has been rescaled by some typical concentration of the problem $\phi_0$, $y$ by $\lambda$ and $x$ by $\lambda/\phi_0^m$.

In Ref. \cite{rusconi08}, the shear-induced diffusion of platelike particles in a flat channel is considered. Two channels meet at the inlet of the main channel, one being charged in particles with concentration $\phi_0$, the other being particle-free. Then this Heaviside-like initial distribution flattens as it is advected along the channel. As the data are restricted to the case where concentrations near the walls haven't changed significantly (that is to say, the observations are made not too far in the channel), the authors suggest to modelize their experiment by the spreading of an Heaviside distribution, with $\phi=\phi_0$ at one side whatever $x$ and $\phi=0$ on the other side. In other words, they consider the lateral wall is  a source of particles.

In that case, it is natural to look for a solution under the form $\phi(x,y)=\Psi(\eta)$, where  $\eta = y x^{-\alpha}$. If such a solution exists, it will mean that there is a broadening of the distribution with a width that grows with $x$ with an exponent $\alpha$.

Another typical diffusion experiment is the spreading of an initial amount of particles initially injected with stationary distribution. In that case, if the width increases as $x^{\alpha}$, particles number conservation implies that the amplitude should decrease as $x^{-\alpha}$. We shall therefore look for solutions  under the form $\phi(x,y)=x^{-\alpha} \Psi(\eta)$. Finally, we include both cases by considering $\phi(x,y)=x^{-\alpha \nu} \Psi(\eta)$, with $\nu=0,1$.

Equation \ref{eq:eqdiff} becomes: \begin{equation}
x^{-1-\alpha \nu}\big(\alpha \nu \Psi + \alpha \eta \Psi'\big)+x^{-\alpha(2+\nu+m\nu)}\big(\Psi^m\Psi'\big)'=0,
\end{equation}

which yields an equation for $\Psi$ as a function of $\eta$ (and not $x$) iff: \begin{equation}
\alpha=\frac{1}{2+m\nu},\label{eq:eqalpha}
\end{equation} in which case there is a  possibility for self-similar $\phi$ with a typical width scaling as $x^{-\alpha}$.

Solutions with reservoirs of particles at one end ($\nu=0$) always spreads with exponent $1/2$ as in the experiment by Rusconi and Stone. On the contrary, the spreading of a fixed quantity of particle depends on the diffusion process considered: the exponent is $1/2$ for Brownian diffusion (and the solution of equation \ref{eq:eqdiff} is the Gaussian profile), but $1/3$ for shear-induced diffusion due to pair interactions. One sees here that the necessity to have particle collision to get diffusion  ($m\ge1$) leads to slower diffusion (subdiffusion), as neighbors are required and the latter are more and more diluted. 3-body interactions will lead to an exponent $1/4$.

Thus, the scalings and shape of solutions strongly depend on initial conditions, a direct consequence of the non-linearity of the advection-diffusion equation.

\section{Analytical solution for $m=1$.}

\begin{figure}
\includegraphics[width=\columnwidth]{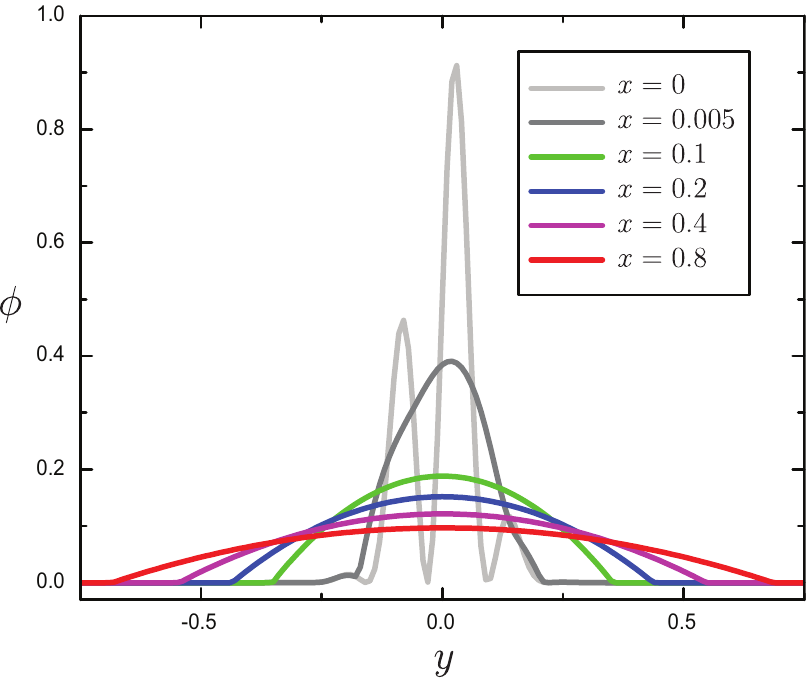}
\caption{\label{simul}(Color online) Evolution of concentration profile along a channel obtained by numerical resolution of equation $\frac{\partial \phi}{\partial x} = \frac{\partial}{\partial y} \big(\phi \frac{\partial \phi}{\partial y}\big)$ with initial condition $\phi(x=0,y)=\frac{1}{2}\big(1+\sin(50 y)\big) \exp\big(-100y^2\big)$.}
\end{figure}

When $\nu=1$, polynomial solutions of Eq. \ref{eq:eqdiff}  can be found, under the form $\Psi(\eta)=0$ or $\Psi(\eta)=-\frac{1}{6}\eta^2+b$, where $b$ is free.  A solution with parabolic profile $\Psi(\eta)=\max(0,-\frac{1}{6}\eta^2+b)$ can then be found. Eq. \ref{eq:eqdiff}  might have other solutions, but numerical resolution with Mathematica software starting from any reasonable or even exotic concentration profile showed convergence towards this self-similar parabolic profile, as illustrated in Fig. \ref{simul}.

Back to the initial units, the half-width at half-height of the concentration profile can be expressed as 
\begin{equation}
w(x)=w_0\big(1+\frac{9 \lambda(y_0) N_0}{4\sqrt{2} w_0^3}x \big)^{1/3},\label{eq:eq2third}
\end{equation}

where  $N_0=\int\phi(x,y)dy$ is a conserved quantity  and $w_0=w(0)$ the initial half-width. At long distance from origin, $w$ does not depend any more on $w_0$. The exponent $1/3$ as well as the dependency of the effective diffusion coefficient on the number of particle $N_0$ are direct evidences for diffusion due to pair interactions.

Note also that the parabolic profile has steeper edges than the Gaussian profile found for Brownian diffusion, which is another indication of the necessity to have neighbors to diffuse. In the case of 3-body interactions ($m=2$),  a solution $\Psi(\eta)=\max(0,\frac{1}{2}\sqrt{b-\eta^2})$ can be found, and the edges are even steeper. The exponent is $1/4$ and the effective diffusion coefficient is proportional to $N_0^2$.

Note that the relative importance of pair and 3-body interactions are controlled by the local concentration, while the widening laws depend on $N_0$, a global variable that can hide different situations: a wide and diluted cloud or a narrow and concentrated one.


\begin{thebibliography}{45}
\expandafter\ifx\csname natexlab\endcsname\relax\def\natexlab#1{#1}\fi
\expandafter\ifx\csname bibnamefont\endcsname\relax
  \def\bibnamefont#1{#1}\fi
\expandafter\ifx\csname bibfnamefont\endcsname\relax
  \def\bibfnamefont#1{#1}\fi
\expandafter\ifx\csname citenamefont\endcsname\relax
  \def\citenamefont#1{#1}\fi
\expandafter\ifx\csname url\endcsname\relax
  \def\url#1{\texttt{#1}}\fi
\expandafter\ifx\csname urlprefix\endcsname\relax\def\urlprefix{URL }\fi
\providecommand{\bibinfo}[2]{#2}
\providecommand{\eprint}[2][]{\url{#2}}

\bibitem[{\citenamefont{Sutera and Skalak}(1993)}]{sutera93}
\bibinfo{author}{\bibfnamefont{S.~P.} \bibnamefont{Sutera}} \bibnamefont{and}
  \bibinfo{author}{\bibfnamefont{R.}~\bibnamefont{Skalak}},
  \bibinfo{journal}{Annu. Rev. Fluid Mech.} \textbf{\bibinfo{volume}{25}},
  \bibinfo{pages}{1} (\bibinfo{year}{1993}).

\bibitem[{poi(1835)}]{poiseuille1835}\bibinfo{author}{\bibfnamefont{J.-M.}\bibnamefont{Poiseuille}}
\bibinfo{journal}{Comptes rendus hebdomadaires des s\'eances de l'Acad\'emie
  des sciences} \textbf{\bibinfo{volume}{1}}, \bibinfo{pages}{554}
  (\bibinfo{year}{1835}).

\bibitem[{\citenamefont{Popel and Johnson}(2005)}]{popel05}
\bibinfo{author}{\bibfnamefont{A.~S.} \bibnamefont{Popel}} \bibnamefont{and}
  \bibinfo{author}{\bibfnamefont{P.~C.} \bibnamefont{Johnson}},
  \bibinfo{journal}{Annu. Rev. Fluid Mech.} \textbf{\bibinfo{volume}{37}},
  \bibinfo{pages}{43} (\bibinfo{year}{2005}).

\bibitem[{\citenamefont{Goldsmith}(1971)}]{goldsmith71}
\bibinfo{author}{\bibfnamefont{H.~L.} \bibnamefont{Goldsmith}},
  \bibinfo{journal}{Fed. Proc.} \textbf{\bibinfo{volume}{30}},
  \bibinfo{pages}{1578} (\bibinfo{year}{1971}).

\bibitem[{\citenamefont{Fung}(1993)}]{fung93}
\bibinfo{author}{\bibfnamefont{Y.~C.} \bibnamefont{Fung}},
  \emph{\bibinfo{title}{Biomechanics: mechanical properties of living tissues}}
  (\bibinfo{publisher}{Springer}, \bibinfo{address}{Berlin},
  \bibinfo{year}{1993}).



\bibitem[{\citenamefont{Bretherton}(1962)}]{bretherton62}
\bibinfo{author}{\bibfnamefont{F.~P.} \bibnamefont{Bretherton}},
  \bibinfo{journal}{J. Fluid Mech.} \textbf{\bibinfo{volume}{14}},
  \bibinfo{pages}{284} (\bibinfo{year}{1962}).

\bibitem[{\citenamefont{Olla}(1997)}]{olla97}
\bibinfo{author}{\bibfnamefont{P.}~\bibnamefont{Olla}}, \bibinfo{journal}{J.
  Phys. II France} \textbf{\bibinfo{volume}{7}}, \bibinfo{pages}{1533}
  (\bibinfo{year}{1997}).

\bibitem[{\citenamefont{Cantat and Misbah}(1999)}]{cantat99}
\bibinfo{author}{\bibfnamefont{I.}~\bibnamefont{Cantat}} \bibnamefont{and}
  \bibinfo{author}{\bibfnamefont{C.}~\bibnamefont{Misbah}},
  \bibinfo{journal}{Phys. Rev. Lett.} \textbf{\bibinfo{volume}{83}},
  \bibinfo{pages}{880} (\bibinfo{year}{1999}).

\bibitem[{\citenamefont{Abkarian et~al.}(2002)\citenamefont{Abkarian, Lartigue,
  and Viallat}}]{abkarian02}
\bibinfo{author}{\bibfnamefont{M.}~\bibnamefont{Abkarian}},
  \bibinfo{author}{\bibfnamefont{C.}~\bibnamefont{Lartigue}}, \bibnamefont{and}
  \bibinfo{author}{\bibfnamefont{A.}~\bibnamefont{Viallat}},
  \bibinfo{journal}{Phys. Rev. Lett.} \textbf{\bibinfo{volume}{88}},
  \bibinfo{pages}{068103} (\bibinfo{year}{2002}).

\bibitem[{\citenamefont{Callens et~al.}(2008)\citenamefont{Callens, Minetti,
  Coupier, Mader, Dubois, Misbah, and Podgorski}}]{callens08}
\bibinfo{author}{\bibfnamefont{N.}~\bibnamefont{Callens}},
  \bibinfo{author}{\bibfnamefont{C.}~\bibnamefont{Minetti}},
  \bibinfo{author}{\bibfnamefont{G.}~\bibnamefont{Coupier}},
  \bibinfo{author}{\bibfnamefont{M.}~\bibnamefont{Mader}},
  \bibinfo{author}{\bibfnamefont{F.}~\bibnamefont{Dubois}},
  \bibinfo{author}{\bibfnamefont{C.}~\bibnamefont{Misbah}}, \bibnamefont{and}
  \bibinfo{author}{\bibfnamefont{T.}~\bibnamefont{Podgorski}},
  \bibinfo{journal}{Europhys. Lett.} \textbf{\bibinfo{volume}{83}},
  \bibinfo{pages}{24002} (\bibinfo{year}{2008}).

\bibitem[{\citenamefont{Goldsmith and Marlow}(1972)}]{goldsmith72}
\bibinfo{author}{\bibfnamefont{H.~L.} \bibnamefont{Goldsmith}}
  \bibnamefont{and} \bibinfo{author}{\bibfnamefont{J.}~\bibnamefont{Marlow}},
  \bibinfo{journal}{Proc. R. Soc. B} \textbf{\bibinfo{volume}{182}},
  \bibinfo{pages}{351} (\bibinfo{year}{1972}).

\bibitem[{\citenamefont{Bagchi}(2007)}]{bagchi07}
\bibinfo{author}{\bibfnamefont{P.}~\bibnamefont{Bagchi}},
  \bibinfo{journal}{Biophys. J.} \textbf{\bibinfo{volume}{92}},
  \bibinfo{pages}{1858} (\bibinfo{year}{2007}).

\bibitem[{\citenamefont{Secomb et~al.}(2007)\citenamefont{Secomb,
  {Styp-Rekowska}, and Pries}}]{secomb07}
\bibinfo{author}{\bibfnamefont{T.~W.} \bibnamefont{Secomb}},
  \bibinfo{author}{\bibfnamefont{B.}~\bibnamefont{{Styp-Rekowska}}},
  \bibnamefont{and} \bibinfo{author}{\bibfnamefont{A.~R.} \bibnamefont{Pries}},
  \bibinfo{journal}{Ann. Biomed. Eng.} \textbf{\bibinfo{volume}{35}},
  \bibinfo{pages}{755} (\bibinfo{year}{2007}).

\bibitem[{\citenamefont{Shi et~al.}(2012)\citenamefont{Shi, Pan, and
  Glowinski}}]{shi10}
\bibinfo{author}{\bibfnamefont{L.}~\bibnamefont{Shi}},
  \bibinfo{author}{\bibfnamefont{T.-W.} \bibnamefont{Pan}}, \bibnamefont{and}
  \bibinfo{author}{\bibfnamefont{R.}~\bibnamefont{Glowinski}},
  \bibinfo{journal}{International Journal for Numerical Methods in Fluids}
  \textbf{\bibinfo{volume}{68}}, \bibinfo{pages}{1393} (\bibinfo{year}{2012}),
  ISSN \bibinfo{issn}{1097-0363}.

\bibitem[{\citenamefont{Li and Ma}(2010)}]{li10}
\bibinfo{author}{\bibfnamefont{H.}~\bibnamefont{Li}} \bibnamefont{and}
  \bibinfo{author}{\bibfnamefont{G.}~\bibnamefont{Ma}}, \bibinfo{journal}{Phys.
  Rev. E} \textbf{\bibinfo{volume}{82}}, \bibinfo{pages}{026304}
  (\bibinfo{year}{2010}).

\bibitem[{\citenamefont{Kumar and Graham}(2012)}]{kumar12}
\bibinfo{author}{\bibfnamefont{A.}~\bibnamefont{Kumar}} \bibnamefont{and}
  \bibinfo{author}{\bibfnamefont{M.~D.} \bibnamefont{Graham}},
  \bibinfo{journal}{Phys. Rev. Lett.} \textbf{\bibinfo{volume}{109}},
  \bibinfo{pages}{108102} (\bibinfo{year}{2012}).

\bibitem[{\citenamefont{Hariprasad and Secomb}(2012)}]{hariprasad12}
\bibinfo{author}{\bibfnamefont{D.~S.} \bibnamefont{Hariprasad}}
  \bibnamefont{and} \bibinfo{author}{\bibfnamefont{T.~W.}
  \bibnamefont{Secomb}}, \bibinfo{journal}{J. Fluid Mech.} \textbf{\bibinfo{volume}{705}},
  \bibinfo{pages}{195}
  (\bibinfo{year}{2012}).

\bibitem[{\citenamefont{Geislinger et~al.}(2012)\citenamefont{Geislinger,
  Eggart, Braunm\"{u}ller, Schmid, and Franke}}]{geislinger12}
\bibinfo{author}{\bibfnamefont{T.~M.} \bibnamefont{Geislinger}},
  \bibinfo{author}{\bibfnamefont{B.}~\bibnamefont{Eggart}},
  \bibinfo{author}{\bibfnamefont{S.}~\bibnamefont{Braunm\"{u}ller}},
  \bibinfo{author}{\bibfnamefont{L.}~\bibnamefont{Schmid}}, \bibnamefont{and}
  \bibinfo{author}{\bibfnamefont{T.}~\bibnamefont{Franke}},
  \bibinfo{journal}{Appl. Phys. Lett.} \textbf{\bibinfo{volume}{100}},
  \bibinfo{pages}{183701} (\bibinfo{year}{2012}).
  
  \bibitem[{\citenamefont{Risser et~al.}(2009)\citenamefont{Risser, Plourabou\'e,
  Cloetens, and Fonta}}]{risser09}
\bibinfo{author}{\bibfnamefont{L.}~\bibnamefont{Risser}},
  \bibinfo{author}{\bibfnamefont{F.}~\bibnamefont{Plourabou\'e}},
  \bibinfo{author}{\bibfnamefont{P.}~\bibnamefont{Cloetens}}, \bibnamefont{and}
  \bibinfo{author}{\bibfnamefont{C.}~\bibnamefont{Fonta}},
  \bibinfo{journal}{Int. J. Dev. Neurosci.} \textbf{\bibinfo{volume}{27}},
  \bibinfo{pages}{185Ð196} (\bibinfo{year}{2009}).

\bibitem[{\citenamefont{Batchelor and Green}(1972)}]{batchelor72a}
\bibinfo{author}{\bibfnamefont{G.~K.} \bibnamefont{Batchelor}}
  \bibnamefont{and} \bibinfo{author}{\bibfnamefont{J.~T.} \bibnamefont{Green}},
  \bibinfo{journal}{J. Fluid Mech.} \textbf{\bibinfo{volume}{56}},
  \bibinfo{pages}{375} (\bibinfo{year}{1972}).

\bibitem[{\citenamefont{Zinchenko}(1984)}]{zinchenko84}
\bibinfo{author}{\bibfnamefont{A.~Z.} \bibnamefont{Zinchenko}},
  \bibinfo{journal}{J. Applied Math. Mech.}
  \textbf{\bibinfo{volume}{48}}, \bibinfo{pages}{198} (\bibinfo{year}{1984}).

\bibitem[{\citenamefont{Da~Cunha and Hinch}(1996)}]{dacunha96}
\bibinfo{author}{\bibfnamefont{F.}~\bibnamefont{Da~Cunha}} \bibnamefont{and}
  \bibinfo{author}{\bibfnamefont{E.}~\bibnamefont{Hinch}}, \bibinfo{journal}{J.
  Fluid Mech.} \textbf{\bibinfo{volume}{309}}, \bibinfo{pages}{211}
  (\bibinfo{year}{1996}).

\bibitem[{\citenamefont{{Van Wijngaarden} and Jeffrey}(1976)}]{wijngaarden76}
\bibinfo{author}{\bibfnamefont{L.}~\bibnamefont{{Van Wijngaarden}}}
  \bibnamefont{and} \bibinfo{author}{\bibfnamefont{D.~J.}
  \bibnamefont{Jeffrey}}, \bibinfo{journal}{J. Fluid Mech.}
  \textbf{\bibinfo{volume}{77}}, \bibinfo{pages}{27} (\bibinfo{year}{1976}).

\bibitem[{\citenamefont{Loewenberg and Hinch}(1997)}]{loewenberg97}
\bibinfo{author}{\bibfnamefont{M.}~\bibnamefont{Loewenberg}} \bibnamefont{and}
  \bibinfo{author}{\bibfnamefont{E.}~\bibnamefont{Hinch}}, \bibinfo{journal}{J.
  Fluid Mech.} \textbf{\bibinfo{volume}{338}}, \bibinfo{pages}{299}
  (\bibinfo{year}{1997}).

\bibitem[{\citenamefont{Lac et~al.}(2007)\citenamefont{Lac, Morel, and
  Barth\`es-Biesel}}]{lac07}
\bibinfo{author}{\bibfnamefont{E.}~\bibnamefont{Lac}},
  \bibinfo{author}{\bibfnamefont{A.}~\bibnamefont{Morel}}, \bibnamefont{and}
  \bibinfo{author}{\bibfnamefont{D.}~\bibnamefont{Barth\`es-Biesel}},
  \bibinfo{journal}{J. Fluid. Mech.} \textbf{\bibinfo{volume}{573}},
  \bibinfo{pages}{149} (\bibinfo{year}{2007}).

\bibitem[{\citenamefont{Kantsler et~al.}(2008)\citenamefont{Kantsler, Segre,
  and Steinberg}}]{kantsler08}
\bibinfo{author}{\bibfnamefont{V.}~\bibnamefont{Kantsler}},
  \bibinfo{author}{\bibfnamefont{E.}~\bibnamefont{Segre}}, \bibnamefont{and}
  \bibinfo{author}{\bibfnamefont{V.}~\bibnamefont{Steinberg}},
  \bibinfo{journal}{Europhys. Lett.} \textbf{\bibinfo{volume}{82}},
  \bibinfo{pages}{58005} (\bibinfo{year}{2008}).

\bibitem[{\citenamefont{Gires et~al.}(2012)\citenamefont{Gires, Danker, and
  Misbah}}]{gires12}
\bibinfo{author}{\bibfnamefont{P.-Y.} \bibnamefont{Gires}},
  \bibinfo{author}{\bibfnamefont{G.}~\bibnamefont{Danker}}, \bibnamefont{and}
  \bibinfo{author}{\bibfnamefont{C.}~\bibnamefont{Misbah}},
  \bibinfo{journal}{Phys. Rev. E} \textbf{\bibinfo{volume}{86}},
  \bibinfo{pages}{011408} (\bibinfo{year}{2012}).



\bibitem[{\citenamefont{Goldsmith and Marlow}(1979)}]{goldsmith79}
\bibinfo{author}{\bibfnamefont{H.~L.} \bibnamefont{Goldsmith}}
  \bibnamefont{and} \bibinfo{author}{\bibfnamefont{J.~C.}
  \bibnamefont{Marlow}}, \bibinfo{journal}{J. Colloid Interface Sci.}
  \textbf{\bibinfo{volume}{71}}, \bibinfo{pages}{383} (\bibinfo{year}{1979}).

\bibitem[{\citenamefont{Cha and Beissinger}(2001)}]{cha01}
\bibinfo{author}{\bibfnamefont{W.}~\bibnamefont{Cha}} \bibnamefont{and}
  \bibinfo{author}{\bibfnamefont{R.~L.} \bibnamefont{Beissinger}},
  \bibinfo{journal}{Korean J. Chem. Eng.} \textbf{\bibinfo{volume}{18}},
  \bibinfo{pages}{479} (\bibinfo{year}{2001}).

\bibitem[{\citenamefont{Higgins et~al.}(2009)\citenamefont{Higgins, Eddington,
  Bhatia, and Mahadevan}}]{Higgins09}
\bibinfo{author}{\bibfnamefont{J.~M.} \bibnamefont{Higgins}},
  \bibinfo{author}{\bibfnamefont{D.~T.} \bibnamefont{Eddington}},
  \bibinfo{author}{\bibfnamefont{S.~N.} \bibnamefont{Bhatia}},
  \bibnamefont{and}
  \bibinfo{author}{\bibfnamefont{L.}~\bibnamefont{Mahadevan}},
  \bibinfo{journal}{PLoS Computational Biology} \textbf{\bibinfo{volume}{5}},
  \bibinfo{pages}{e1000288} (\bibinfo{year}{2009}).

\bibitem[{\citenamefont{Zhao and Shaqfeh}(2011)}]{zhao11}
\bibinfo{author}{\bibfnamefont{H.}~\bibnamefont{Zhao}} \bibnamefont{and}
  \bibinfo{author}{\bibfnamefont{E.~S.~G.} \bibnamefont{Shaqfeh}},
  \bibinfo{journal}{Phys. Rev. E} \textbf{\bibinfo{volume}{83}},
  \bibinfo{pages}{061924} (\bibinfo{year}{2011}).

\bibitem[{\citenamefont{Tan et~al.}(2012)\citenamefont{Tan, Le, and
  Chiam}}]{tan12}
\bibinfo{author}{\bibfnamefont{M.~H.-Y.} \bibnamefont{Tan}},
  \bibinfo{author}{\bibfnamefont{D.-V.} \bibnamefont{Le}}, \bibnamefont{and}
  \bibinfo{author}{\bibfnamefont{K.-H.} \bibnamefont{Chiam}},
  \bibinfo{journal}{Soft Matter} \textbf{\bibinfo{volume}{8}},
  \bibinfo{pages}{2243} (\bibinfo{year}{2012}).

\bibitem[{\citenamefont{Hudson}(2003)}]{hudson03}
\bibinfo{author}{\bibfnamefont{S.~D.} \bibnamefont{Hudson}},
  \bibinfo{journal}{Physics of Fluids} \textbf{\bibinfo{volume}{15}},
  \bibinfo{pages}{1106} (\bibinfo{year}{2003}).

\bibitem[{\citenamefont{Rusconi and Stone}(2008)}]{rusconi08}
\bibinfo{author}{\bibfnamefont{R.}~\bibnamefont{Rusconi}} \bibnamefont{and}
  \bibinfo{author}{\bibfnamefont{H.~A.} \bibnamefont{Stone}},
  \bibinfo{journal}{Phys. Rev.Lett.} \textbf{\bibinfo{volume}{101}},
  \bibinfo{pages}{254502} (\bibinfo{year}{2008}).

\bibitem[{\citenamefont{Pranay et~al.}(2012)\citenamefont{Pranay,
  {Henriquez-Rivera}, and Graham}}]{pranay12}
\bibinfo{author}{\bibfnamefont{P.}~\bibnamefont{Pranay}},
  \bibinfo{author}{\bibfnamefont{R.~G.} \bibnamefont{{Henriquez-Rivera}}},
  \bibnamefont{and} \bibinfo{author}{\bibfnamefont{M.~D.}
  \bibnamefont{Graham}}, \bibinfo{journal}{Phys. Fluids}
  \textbf{\bibinfo{volume}{24}}, \bibinfo{pages}{061902}
  (\bibinfo{year}{2012}).

\bibitem[{\citenamefont{Dubois et~al.}(2006{\natexlab{a}})\citenamefont{Dubois,
  Callens, Yourassowsky, Hoyos, Kurowski, and Monnom}}]{Dubois06_2}
\bibinfo{author}{\bibfnamefont{F.}~\bibnamefont{Dubois}},
  \bibinfo{author}{\bibfnamefont{N.}~\bibnamefont{Callens}},
  \bibinfo{author}{\bibfnamefont{C.}~\bibnamefont{Yourassowsky}},
  \bibinfo{author}{\bibfnamefont{M.}~\bibnamefont{Hoyos}},
  \bibinfo{author}{\bibfnamefont{P.}~\bibnamefont{Kurowski}}, \bibnamefont{and}
  \bibinfo{author}{\bibfnamefont{O.}~\bibnamefont{Monnom}},
  \bibinfo{journal}{Appl. Opt.} \textbf{\bibinfo{volume}{45}},
  \bibinfo{pages}{864} (\bibinfo{year}{2006}{\natexlab{a}}).

\bibitem[{\citenamefont{Dubois et~al.}(2006{\natexlab{b}})\citenamefont{Dubois,
  Yourassowsky, Monnom, Legros, Debeir, Van~Ham, Kiss, and
  Decaestecker}}]{dubois06_1}
\bibinfo{author}{\bibfnamefont{F.}~\bibnamefont{Dubois}},
  \bibinfo{author}{\bibfnamefont{C.}~\bibnamefont{Yourassowsky}},
  \bibinfo{author}{\bibfnamefont{O.}~\bibnamefont{Monnom}},
  \bibinfo{author}{\bibfnamefont{J.-C.} \bibnamefont{Legros}},
  \bibinfo{author}{\bibfnamefont{O.}~\bibnamefont{Debeir}},
  \bibinfo{author}{\bibfnamefont{P.}~\bibnamefont{Van~Ham}},
  \bibinfo{author}{\bibfnamefont{R.}~\bibnamefont{Kiss}}, \bibnamefont{and}
  \bibinfo{author}{\bibfnamefont{C.}~\bibnamefont{Decaestecker}},
  \bibinfo{journal}{J. Biomed. Opt.}
  \textbf{\bibinfo{volume}{11}}, \bibinfo{pages}{054032}
  (\bibinfo{year}{2006}{\natexlab{b}}).


\bibitem[{\citenamefont{Koter}(1990)}]{koter90}
\bibinfo{author}{\bibfnamefont{M.}~\bibnamefont{Koter}},
  \bibinfo{journal}{Int. J. Radiat. Biol.}
  \textbf{\bibinfo{volume}{58}}, \bibinfo{pages}{157} (\bibinfo{year}{1990}).

\bibitem[{\citenamefont{Fischer}(2007)}]{fisher07}
\bibinfo{author}{\bibfnamefont{T.~M.} \bibnamefont{Fischer}},
  \bibinfo{journal}{Biophys J.} \textbf{\bibinfo{volume}{93}},
  \bibinfo{pages}{2553} (\bibinfo{year}{2007}).
  
  \bibitem[{\citenamefont{Turgeon}(2005)}]{turgeon05}
\bibinfo{author}{\bibfnamefont{M.~L.} \bibnamefont{Turgeon}},
  \emph{\bibinfo{title}{Clinical Hematology: Theory and Procedures.}}
  (\bibinfo{publisher}{Lippincott Williams \& Wilkins}, \bibinfo{address}{Philadelphia, USA},
  \bibinfo{year}{2005}).


\bibitem[{sup()}]{suppl}
\bibinfo{note}{See supplemental material}.

\bibitem[{\citenamefont{White}(1991)}]{white}
\bibinfo{author}{\bibfnamefont{F.~M.} \bibnamefont{White}},
  \emph{\bibinfo{title}{Viscous Fluid Flow}} (\bibinfo{publisher}{McGraw-Hill},
  \bibinfo{address}{New York}, \bibinfo{year}{1991}).

\bibitem[{\citenamefont{Abkarian et~al.}(2007)\citenamefont{Abkarian, Faivre,
  and Viallat}}]{abkarian07}
\bibinfo{author}{\bibfnamefont{M.}~\bibnamefont{Abkarian}},
  \bibinfo{author}{\bibfnamefont{M.}~\bibnamefont{Faivre}}, \bibnamefont{and}
  \bibinfo{author}{\bibfnamefont{A.}~\bibnamefont{Viallat}},
  \bibinfo{journal}{Phys. Rev. Lett.} \textbf{\bibinfo{volume}{98}},
  \bibinfo{pages}{188302} (\bibinfo{year}{2007}).

\bibitem[{\citenamefont{Dupire et~al.}(2010)\citenamefont{Dupire, Abkarian, and
  Viallat}}]{dupire10}
\bibinfo{author}{\bibfnamefont{J.}~\bibnamefont{Dupire}},
  \bibinfo{author}{\bibfnamefont{M.}~\bibnamefont{Abkarian}}, \bibnamefont{and}
  \bibinfo{author}{\bibfnamefont{A.}~\bibnamefont{Viallat}},
  \bibinfo{journal}{Phys. Rev. Lett.} \textbf{\bibinfo{volume}{104}},
  \bibinfo{pages}{168101} (\bibinfo{year}{2010}).

\bibitem[{\citenamefont{Bishop et~al.}(2002)\citenamefont{Bishop, Popel,
  Intaglietta, and Johnson}}]{Bishop02}
\bibinfo{author}{\bibfnamefont{J.~J.} \bibnamefont{Bishop}},
  \bibinfo{author}{\bibfnamefont{A.~S.} \bibnamefont{Popel}},
  \bibinfo{author}{\bibfnamefont{M.}~\bibnamefont{Intaglietta}},
  \bibnamefont{and} \bibinfo{author}{\bibfnamefont{P.~C.}
  \bibnamefont{Johnson}}, \bibinfo{journal}{Am. J. Physiol. Heart Circ.
  Physiol.} \textbf{\bibinfo{volume}{283}}, \bibinfo{pages}{H1985}
  (\bibinfo{year}{2002}).

\end{thebibliography}
\end{document}